\input harvmac

\def\np#1#2#3{Nucl. Phys. {\bf B#1} (#2) #3}
\def\pl#1#2#3{Phys. Lett. {\bf #1B} (#2) #3}
\def\prl#1#2#3{Phys. Rev. Lett. {\bf #1} (#2) #3}

\def\cl{{\cal L}}

\Title{hep-th/9512145, EFI-95-79}
{\vbox{\centerline{Orbifolds and Solitons}
}}
\bigskip
\centerline{D. Kutasov}
\vglue .5cm
\centerline{Enrico Fermi Institute and}
\centerline{Department of Physics}
\centerline{University of Chicago}
\centerline{Chicago, IL 60637, USA}
\vglue .3cm

\bigskip\bigskip

\noindent
We propose a conformal field theory
description of a solitonic heterotic string in
type $IIA$ superstring theory compactified
on $K3$, generalizing previous work by
J. Harvey, A. Strominger and A. Sen.
In ten dimensions the construction gives
a fivebrane which is related to the fundamental
type $II$ string by electric -- magnetic duality,
and to the Dirichlet fivebrane of type $IIB$ string theory
by $SL(2, Z)$.

\Date{12/95}

Recently, important progress has been made in analyzing
non-perturbative properties of various string theories
using string duality (see
\ref\polc{J. Polchinski, hep-th/9511157.}\
for a review and a list of references). Among the most
remarkable results is the discovery by J. Polchinski of a simple
description of extended objects (Dirichlet-branes) carrying various
Ramond -- Ramond (RR) charges
\ref\polcc{J. Polchinski, hep-th/9510017, \prl{75}{1995}{4724}.}.
One of the many applications of D-branes has been in providing
evidence for string duality by constructing conformal field
theories (CFT's) corresponding to dual strings.

Thus, for example, the D-string of the type $I$ theory
in ten dimensions has been shown \ref\pw{J. Polchinski
and E. Witten, hep-th/9510169, \np{460}{1996}{525}.}\ to possess the massless
world sheet degrees of freedom of the heterotic string
with gauge group $SO(32)$ (see also
\nref\atish{A. Dabholkar, hep-th/9506160, \pl{357}{1995}{307}.}%
\nref\hull{C. Hull, hep-th/9506194, \pl{357}{1995}{545}.}%
\refs{\atish, \hull}),
while the D-string of the
$IIB$ theory in ten dimensions has been found to describe the
$SL(2, Z)$ dual of the original fundamental string
\ref\wit{E. Witten, hep-th/9510135, \np{460}{1996}{335}.}.
In both cases one learns by matching
the low energy effective actions that the dual string
should have tension proportional to the inverse of the
fundamental string coupling and should couple to the
Ramond -- Ramond $B_{\mu\nu}$ field of the fundamental
string, in agreement with properties of D-strings \polcc.

\nref\ew{E. Witten, hep-th/9503124, \np{443}{1995}{85}.}%
\nref\nati{N. Seiberg, Nucl. Phys. {\bf B303} (1988) 286.}%
\nref\aspin{P. Aspinwall and D. Morrison, hep-th/9404151.}%
\nref\hullt{C. Hull and P. Townsend, hep-th/9410167,
\np{438}{1995}{109}.}%
\nref\duff{M. Duff, hep-th/9501030, \np{442}{1995}{47};
M. Duff and and R. Khuri, hep-th/9305142, \np{411}{1994}{473}.}%
\nref\jh{J. Harvey and A. Strominger, hep-th/9504047,
\np{449}{1995}{535}.}%
\nref\as{A. Sen, hep-th/9504027, \np{450}{1995}{103}.}%
In the general discussion of string duality (see e.g.
\refs{\polc, \ew}) a central
role is played by the conjectured equivalence of the
type $IIA$ string compactified to six dimensions on $K3$,
and the heterotic string compactified on $T^4$. This
equivalence determines the
strong coupling dynamics of the heterotic string
in all dimensions between four (where it leads
to $S$ duality) and seven.
There is a lot of evidence for this symmetry
\refs{\ew  - \as},
including a construction of the heterotic string as a solitonic
solution of the low energy equations of motion of the type $IIA$
string on $K3$ \refs{\jh, \as}.
The purpose of this note is to propose a CFT description of
that construction.

There are a number of (equivalent)
ways to see that the solitonic heterotic string of \refs{\jh, \as}
is not a D-string.
The $IIA$ theory on $K3$ does not have Ramond --
Ramond $B_{\mu\nu}$ fields. Hence it can not couple to
D-strings. Also, by matching the low energy effective
actions of the $IIA$ string on $K3$ and heterotic string
on $T^4$ one finds \ew\ that the tension of the solitonic
heterotic string should scale like the {\it square}
of the inverse string coupling of the $IIA$ theory,
suggesting that one should
look for a construction of the solitonic string as a closed
string, which gets its tension from the sphere (rather than
the disk as in the case of D-strings \polcc). The Kalb -- Ramond
field that the solitonic heterotic string couples to is the six
dimensional Poincare dual of
the NS -- NS Kalb -- Ramond field of the type $IIA$ string.

We will argue that the CFT describing a solitonic
heterotic string corresponds to the type $IIA$ string
propagating in the background
\eqn\backg{R^2\times R^4/I_2\times K3,}
where
$R^2$ is the world sheet of the solitonic string, and
$R^4/I_2$ is a non compact $Z_2$ orbifold with the string sitting at the
fixed point. 
The $Z_2$ symmetry $I_2$ changes the sign of all four coordinates
on $R^4$ and acts as $(-)^{F_L}$ on the left moving fermions. We will show
that the twisted sector modes living at the fixed point  of $R^4/I_2$ 
describe the world sheet degrees of freedom of a heterotic string, 
propagating on $R^2\times R^4/Z_2\times T^4$. 
This description is presumably related
to that of \refs{\jh, \as} in the same way as the type $I$ D-string
of \pw\ is related to the construction of
\refs{\atish, \hull}. In ten dimensions (i.e. replacing $K3$
by $R^4$ in \backg) the same construction gives a fivebrane
which apparently couples to the six form dual of the NS -- NS
$B_{\mu\nu}$ field of type $II$ string theory, completing the list
of extended objects that carry charges of the various $p$ -- form
gauge fields in type $II$ theories.

While the construction presented here is general, for concreteness
and simplicity we will describe it for the case when the $K3$
surface is an orbifold, $T^4/Z_2$. This will also allow us to make
a certain point later.
It is easy to generalize the discussion to an arbitrary $K3$ surface.
We start by reviewing some aspects of the structure of type $II$
string theory on $R^6\times T^4/Z_2$.
For later convenience we describe $R^6$ as $R^2\times R^4$.
The coordinates on the space-time $R^2\times R^4\times T^4/Z_2$
will be denoted by:
\eqn\cord{(x^\pm, y^\pm_\mu, w^\pm_m);\;\;\mu=1,2,;\; m=3,4.}
The coordinates $\{x^\pm, y^\pm_\mu\}$ parametrize $R^6$;
$\{w^\pm_m\}$ live in $T^4/Z_2$.
We will also need the world sheet fermions $(\psi^\pm, \psi^\pm_\mu,
\psi^\pm_m)$.
After bosonization, $\{\psi^\pm\}$ give rise to a single scalar, $H$;
$\{\psi^\pm_\mu\}$ and $\{\psi^\pm_m\}$
give two scalars each, $(H_1, H_2)$
and $(H_3, H_4)$, respectively.
The right moving superconformal generator
is given by:
\eqn\tfsc{T_F=\psi\cdot\partial X=e^{iH}\partial x^+ +
e^{-iH}\partial x^-+e^{iH_\mu}\partial
y^+_\mu+e^{-iH_\mu}\partial y^-_\mu+
e^{iH_m}\partial w^+_m+e^{-iH_m}\partial w^-_m}
where $\mu=1,2$; $m=3,4$.
Similar formulae hold for the left movers.

In ten dimensions type $II$ theories have 32 supercharges. The
$T^4/Z_2$ ($K3$) background leaves 16 unbroken. In the $IIA$
theory they are:
\eqn\superch{\eqalign{
Q_\alpha^\pm=&\oint dz e^{-{\phi\over2}}S_\alpha
e^{\pm{i\over2}(H_3+H_4)}\cr
\bar Q^\pm_{\bar\alpha}=&\oint d\bar z e^{-{\bar\phi\over2}}
\bar S_{\bar \alpha}e^{\pm{i\over2}(\bar H_3+\bar H_4)}\cr}}
As usual, $S_\alpha$ is a dimension $3/8$ spin field in
the $4$ of $SO(6)$ and $\phi$, $\bar\phi$ are bosonized ghost fields
\ref\fms{D. Friedan, E. Martinec and S. Shenker, \np{271}{1986}{93}.}.
The right moving supercharges belong to
the $4$ of $SO(6)$. The left moving
ones belong to a $\bar 4$.
Decomposing the supercharges \superch\ under
$SO(2)\times SO(4)\times SO(4)$ which acts on the three factors in
\cord\ we find that the supercharges transform as follows:
\eqn\trn{\eqalign{
Q:&\;\;(2^+, 2^+)_+\; +\;(2^-, 2^+)_-\cr
\bar Q:&\;\;(2^+, 2^+)_-\;+\;(2^-, 2^+)_+\cr}
}
The superscripts in \trn\ refer to the $SO(4)$ chirality; the subscripts
to $SO(2)$ chirality.
In the untwisted sector of the $T^4/Z_2$ orbifold we project
onto states even under $(w_m,\psi_m)\to-(w_m, \psi_m)$. The twisted
sectors are created by applying the twist fields $\sigma_i$, $i=1,
\cdots, 16$ to the vacuum. The fields $\sigma_i$ which have conformal
dimension $(1/4, 1/4)$ are in one to one correspondence with the 16
fixed points of the $Z_2$ symmetry $w\to-w$. The $K3$ CFT has 80
moduli parametrizing the symmetric space
$SO(20,4)/SO(20)\times SO(4)$ \nati. In the $T^4/Z_2$ CFT
16 of those arise as the vertex operators (in the $-1$ picture \fms)
$\exp(-\phi-\bar\phi)\psi_m\bar\psi_n$, $m,n=1,\cdots,4$, controlling
the size and shape of the four torus, while
the other $64=4\times 16$ are blowing up
modes associated with the fixed points:
$$e^{-\phi-\bar\phi}\sigma_ie^{\pm{i\over2}(H_3+H_4)}
e^{\pm{i\over2}(\bar H_3+\bar H_4)}.$$ 
The type $IIA$ theory
has also 24 $U(1)$ gauge fields in the RR sector. Sixteen are given
by:
\eqn\ga{F_{\mu\nu}^{(i)}e^{-{\phi\over2}-{\bar\phi\over2}}
\gamma^{\mu\nu}_{\bar\alpha\beta}\sigma_iS_{\bar\alpha}\bar S_\beta.}
The other eight:
\eqn\gb{\eqalign{
&F_{\mu\nu}^{(\pm, \pm)}e^{-{\phi\over2}-{\bar\phi\over2}}
\gamma^{\mu\nu}_{\bar\alpha\beta} e^{\pm{i\over2}(H_3-H_4)}
e^{\pm{i\over2}(\bar H_3-\bar H_4)}S_{\bar\alpha}\bar S_\beta\cr
&G_{\mu\nu}^{(\pm, \pm)}e^{-{\phi\over2}-{\bar\phi\over2}}
\gamma^{\mu\nu}_{\alpha\bar\beta} e^{\pm{i\over2}(H_3+H_4)}
e^{\pm{i\over2}(\bar H_3+\bar H_4)}S_\alpha\bar S_{\bar\beta}\cr
}}

Next we consider this model on $R^2\times R^4/I_2\times T^4/Z_2$
(compare to \backg), i.e. modd out by the symmetry $y\to-y$ and $(-)^{F_L}$ 
in \cord. This introduces a conical singularity at $y=0$, which
we will intrepret as the location of the soliton string.
Most of the discussion of $T^4/Z_2$ above applies to the non-compact
orbifold $R^4/I_2$. The only slightly subtle issue is the structure
of the twisted sector\foot{I thank J. Harvey and A. Schwimmer
for discussions of this issue.}. Physically one expects a unique
twist field $\Sigma$ with conformal dimension $(1/4, 1/4)$,
since there is a unique fixed point of the $Z_2$
symmetry. On the other hand, if we
define the CFT on $R^4/I_2$ by a large size limit of a $T^4/I_2$,
it appears that modular invariance requires the presence of $16$
twist fields, just as in the discussion of the $K3$ orbifold
above (see
\ref\jgjh{J. Gauntlett and J. Harvey, hep-th/9407111.}\
for a discussion of this issue).
We expect the right definition to be the one suggested by
geometry: as the size of a $T^4/Z_2$ goes to infinity the
$16$ twisted sectors  decouple. It is consistent to keep any
one of them.

The six dimensional background describes a flat space with a conical
singularity at a point on $R^4$. This is a plane in $R^2\times R^4/Z_2$
which we will interpret as the world sheet of the soliton string.
To understand the dynamics of this object we must analyze the collective
modes
propagating on this world sheet. In other words we are looking
for massless string states which include a factor of the twist field
$\Sigma$, and therefore
live at the fixed point of $I_2$.

Just like the $K3$ background breaks half of the 32 supercharges
of the ten dimensional theory, the $R^4/I_2$ background breaks
half of the remaining ones (it is a BPS state). Indeed, the presence
of 16 RR ground states with vertex operators $(i=1,\cdots, 16)$:
\eqn\sixteen{V_i^{(l)}=\int d^2z V_i^{(l)}(x^\pm)
e^{-{\phi\over2}-{\bar\phi\over2}}\Sigma\sigma_i e^{-{i\over2}H-{i\over2}
\bar H} }
implies that the unbroken supercharges are
chiral on the soliton string world sheet $(x^+, x^-)$, transforming
as $(2^-, 2^+)_-$ and $(2^+, 2^+)_-$ (compare to \superch, \trn).
The algebra of the unbroken supercharges is:
\eqn\sch{\{Q^a, Q^b\}=\delta^{ab}P^+;\;a,b=1,\cdots,8.}
We also see that the unbroken supercharges commute with
$V_i^{(l)}$. This means that the $V_i^{(l)}$ are left moving
on the ``world sheet'' $(x^+, x^-)$ (justifying the
superscripts in \sixteen). Indeed, applying the
BRST charge $Q_{BRST}=\oint\gamma T_f+\cdots$ and using \tfsc\
we find that BRST invariance of \sixteen\ implies the equation
of motion:
\eqn\eqmt{{\partial\over\partial x^-} V_i^{(l)}=0;\; i=1,\cdots, 16.}

Additional scalars on the world sheet arise as the four
RR states:
\eqn\lfour{V_{\pm,\pm}^{(l)}=
\int d^2z V_{\pm,\pm}^{(l)}(x^+) e^{-{\phi\over2}-
{\bar\phi\over2}}\Sigma e^{\pm{i\over2}(H_3-H_4)}
e^{\pm{i\over2}(\bar H_3-\bar H_4)}e^{-{i\over 2}H}e^{-{i\over2}\bar H}}
which again commute with the supercharges and are 
left moving on the world sheet as indicated, and the RR states
\eqn\rfour{V_{\pm,\pm}^{(r)}=
\int d^2z V_{\pm,\pm}^{(r)}(x^-) e^{-{\phi\over2}-
{\bar\phi\over2}}\Sigma e^{\pm{i\over2}(H_3+H_4)}
e^{\pm{i\over2}(\bar H_3+\bar H_4)}e^{ {i\over 2}H}e^{ {i\over2}\bar H}}
Applying the unbroken supercharges to the right moving
scalars \rfour\ gives rise to
four right moving fermions on the world sheet.

Thus the RR sector gives four right moving \rfour\ and
twenty left moving \sixteen, \lfour\ chiral scalar fields,
$(V_I^{(r)}(x^+), V_A^{(l)}(x^-))$, $I=1,\cdots, 4$, $A=1, \cdots, 20$
on the soliton string world sheet. The $SO(20, 4; Z)$ discrete
symmetry of the $K3$ CFT \aspin\ apparently implies that
these RR fields live on an even self dual Lorentzian
lattice, $\Lambda^{20,4}$, which changes as we vary the eighty
moduli of $K3$ mentioned above. The world sheet Lagrangian
for these scalars is
\eqn\leff{\cl={1\over\lambda^2\alpha^\prime}\left(
\partial_+ V_I^{(r)}\partial_- V_I^{(r)}+
\partial_+ V_A^{(l)}\partial_- V_A^{(l)}\right)}
where $\lambda$ is the type $IIA$ string coupling, and we have
absorbed in $1/\alpha^\prime$ a factor of the volume
of the $K3$ surface. One may wonder
why the RR fields $V_A, V_I$ have a factor of $1/\lambda^2$
in their kinetic terms \leff\ -- usually RR Lagrangians
are written without such factors. Here \leff\ is correct because
in other discussions RR fields are redefined from
\sixteen, \lfour, \rfour\ by a factor of $\lambda$
to make certain gauge transformations natural (see e.g. footnote
2 of \polcc). Such a redefinition is unnecessary here.
Eq. \leff\ implies that the soliton string tension is
$1/\lambda^2\alpha^\prime$, in agreement with \refs{\jh, \as}\
and string duality.

To complete the picture we consider the massless modes in the
(NS,NS) sector.
These are described by the vertex operators:
\eqn\ffy{Y_{\pm,\pm}=\int d^2z Y_{\pm,\pm}(x^+, x^-)
e^{-\phi-\bar\phi}\Sigma
e^{\pm{i\over2}(H_1-H_2)\pm{i\over2}(\bar H_1+\bar H_2)}}
We find four right moving and four left moving massless scalars in the
vector representation of the $SO(4)$ which acts on the $y^\pm_\mu$
in \cord. They can be thought of as describing
collective transverse fluctuations of the soliton string in $R^4/Z_2$.
Acting on the four right movers in \ffy\ with the supercharges
produces four massless fermions on the worldsheet.

Note the relative minus sign between the left and right moving
$H_i$ in \ffy. This is a choice of GSO projection.
Normally, one takes this relative sign to be plus, thus obtaining
the blowing up modes of the orbifold (see the discussion after eq.
\trn). These transform as $2\times 2=3+1$
under $SO(4)$, corresponding to three classical
blowing up modes and a world sheet theta angle.
The other choice, taken in \ffy\ to account for the
fact that $I_2$ contains a $(-)^{F_L}$ transformation, 
leads to these modes transforming
as $2\times \bar2$ of $SO(4)$ and describing
collective motions of the solitonic strings (or fivebranes) located
at the orbifold fixed points. For a compact manifold
(e.g. replacing $R^4$ by $T^4$), $T$ duality leads to additional
equivalences. For example, a type $IIA$ orbifold with blowing up
modes is mapped by $T$ duality to
a type $IIB$ one with the blowing up modes reinterpreted
as describing collective transverse motions of a fivebrane
(see comment 5 below).

Summarizing, we found that the type $IIA$ string propagating on the
manifold \backg\ with an additional $(-)^{F_L}$ 
projection describes a space-time with a soliton string whose world sheet
is sitting at the conical singularity on $R^4/Z_2$; the modes propagating
on that world sheet were shown to be described by the world sheet
Lagrangian of a heterotic string on $R^2\times R^4/Z_2\times T^4$. As in
\refs{\jh, \as} it is interesting that a chiral structure
arises from the non -- chiral type $IIA$ string. Here it is due to
the chiral $(-)^{F_L}$ projection.

A few comments are in order:

\item {1.} The construction could have been presented
directly in ten dimensions.
There it gives rise to the symmetric fivebrane solution of
\ref\chs{C. Callan, J. Harvey and A. Strominger, \np{359}{1991}{611};
{\bf B367} (1991) 60.}
represented as the orbifold $R^6\times R^4/I_2$. The massless
collective modes on the world volume of the fivebrane arising from
the NS-NS sector are as before four massless scalars $Y_\mu(x^m)$
($\mu=6,\cdots, 9$, $m=0,\cdots, 5$) corresponding to fluctuations of
the fivebrane in $R^4/Z_2$. In the RR sector we find:
1) for the $IIB$ string: a vector $F_{mn}(x^m)\gamma^{mn}_{\alpha\bar\beta}
\exp(-\phi/2-\bar\phi/2)\Sigma S^\alpha\bar S^{\bar\beta}$.
2) for the $IIA$ string: a scalar, and a self dual two form.
As is clear from this matter content, the sixteen unbroken
supercharges correspond to $(1,1)$ supersymmetry on the world volume
for the type $IIB$ theory (i.e. the supercharges form two $4$'s
and two $\bar 4$'s of $SO(6)$), and $(2,0)$ supersymmetry on the world
volume in the $IIA$ case. Due to the chiral $(-)^{F_L}$ projection 
the non-chiral $IIA$ theory
gives rise to a chiral fivebrane 
world volume action and vice versa for $IIB$.
All this is in agreement with the structure found in \chs. For the
type $IIB$ theory it is also in agreement with properties of the $D$-fivebrane
related to the NS-NS fivebrane discussed here by the strong -- weak
coupling $SL(2, Z)$ symmetry.
The solitonic
string in six dimensions discussed above corresponds
to a type $IIA$ fivebrane wrapped around $K3$.

\item {2.} It is interesting to think about the relation
of the world sheet CFT description of the symmetric fivebrane
in \chs\ and the one given here. An important difference
is that while the construction of \chs\ describes a fivebrane
embedded in $R^{10}$, we found that a particularly
simple description emerges if the fivebrane lives in
$R^6\times R^4/Z_2$, and furthermore resides at the
$Z_2$ fixed point. Both the fivebrane and the fixed point
(which can be thought of as an ``orientifold plane'')
carry magnetic $B_{\mu\nu}$ charge; when they coincide
their charges cancel each other such that the total charge
is zero. In \chs\ the description
simplifies in the limit $y^2\to 0$ (the ``throat''
approximation) where one finds an exact
coset construction of the fivebrane. However, it is not clear
how to connect the coset description to flat space as $y^2\to \infty$.
Our construction focuses on the opposite
limit; the `throat' is squeezed to a point $y=0$
and space is flat everywhere else. It is conceivable that if one
perturbs the throat coset CFT in order to connect to flat space
(which is clearly a singular perturbation), the IR limit of the
resulting theory becomes equivalent to the orbifold described here
(i.e. the throat effectively shrinks to zero size).
In \chs\ it was also argued that the solution of the low energy equations
of motion is not corrected and gives rise to an exact CFT, due to
the $(4,4)$ supersymmetry structure. However, as pointed out in \chs\
outside the ``throat'' approximation it is difficult to study
that CFT explicitly. It would be interesting to compare it to the
orbifold in detail.

\item {3.} One can ask whether the orbifold construction
yields other extended objects of different dimensions. This
would seem undesirable since the fivebrane in ten dimensions
and string in six dimensions we constructed have a natural
interpretation as objects that couple magnetically 
to the NS-NS $B_{\mu\nu}$
field. Other $p$ -- branes would not couple
to known massless $p+1$ form gauge potentials. Indeed, as is clear
from the details of our construction, the fivebrane in ten
dimensions (or string in six, etc) is very special. For other
$p$ -- branes constructed using non compact orbifolds, we would
generically find no massless modes on the brane. These $p$-branes
would thus be rigid and presumably not physically interesting.

\item {4.} To describe $n$ heterotic string configurations
one may introduce $n$ twist operators $\Sigma_i$, $i=1, \cdots, n$
corresponding to the $n$ strings (or fivebranes in ten dimensions).
This describes $n$ well separated solitonic strings. To move
them one may add appropriate translational modes $Y^i_\mu$ \ffy\
to the world sheet Lagrangian of the fundamental string.

\item {5.} One can use the basic idea of identifying orbifolds
with solitons directly on $R^6\times K^3$. Consider a compactification
of the $IIA$ string on a $K3$ of the form $T^4/Z_2$ (the
case described above). Performing a $T$ duality transformation
on a single circle 
we can use our results
to think of this as a compactification of
the type
$IIB$ string with $16$ fivebranes distributed 
uniformly on the $T^4/Z_2$.
Then, performing an $SL(2, Z)$ transformation (which exchanges
NS-NS with RR and in particular the NS-NS fivebrane with the
Dirichlet fivebrane) we find a type $IIB$ string compactified
on a ``D manifold'' \wit,
\ref\bsv{M. Bershadsky, V. Sadov and C. Vafa,
hep-th/9510225.}, a four torus with 16 Dirichlet fivebranes and
16 orientifold planes representing the fixed points of $T^4/Z_2$.
The moduli of the original $K3$ we started with are translated
in that theory into $16$ moduli deforming the shape of
the $T^4/Z_2$ and $64$ correspoding to translations of the
$D$-branes. This point of view\foot{See 
\ref\ssss{A. Sen, hep-th/9604070.}\ for additional discussion
of these issues.}
leads to a simple picture of the relation between
type $IIA$ strings compactified on $K3$ and type $IIB$
strings on D-manifolds. The key observation 
is that $T$ duality 
turns blowing up modes of a $T^4/Z_2$
orbifold into moduli controlling the positions of
fivebranes on $T^4/Z_2$. 

\item{6.} The resulting picture can also be used to study qualitative
aspects of the behavior of type $II$ theories on $K3$ surfaces, in 
particular the singular limits that received a lot of attention
in the last year \ew, \ref\hhhh{C. Hull, hep-th/9512181.}. 
It makes it clear that an orbifold $K3$ does not
correspond to a singular CFT since it is equivalent (after T duality
on one of the circles) to a non-singular configuration of fivebranes.
To get a singular CFT one must turn on twisted sector moduli 
\ref\aaas{P. Aspinwall, hep-th/9507012, \pl{357}{1995}{329}.}\
whose effect in the T-dual language is to translate the fivebranes.
A singular CFT arises when two or more of the fivebranes coincide.
The nature of the singularity differs in the type $IIA$ and $IIB$
theories\foot{
See \ref\usss{D. Kutasov, E. Martinec and M. O'Loughlin, hep-th/9603116.}\
for further discussion.}.
Type $IIA$ on a singular $K3$ is equivalent to type $IIB$ on $T^4/Z_2$
with nearly coincident fivebranes. Dirichlet strings
stretched between the two fivebranes become light in the limit, and give rise
to an enhanced gauge symmetry \ew. Type $IIB$ on a singular $K3$
can be described as a $IIA$ theory with coincident fivebranes. This time
the objects that can end on the NS-NS fivebranes are Dirichlet two branes
\ref\sttr{A. Strominger, hep-th/9512059.}.
When the fivebranes coincide, one finds tensionless strings embedded
in the fivebrane \ref\eddd{E. Witten, hep-th/9507121.}. 

There are many interesting open questions regarding the objects
described here. It would be interesting to construct a useful
description of many fivebrane (or string) configurations. Such
configurations seem to play an important role in duality \wit.
An important open question is how to construct the type $IIA$
string as a soliton in the heterotic string theory, or equivalently
a fivebrane in ten dimensional heterotic string theory
\ref\ewit{E. Witten, hep-th/9511030, \np{460}{1996}{541}.}.
These and other questions will be left for future work.

\bigskip

\centerline{{\bf Acknowledgments}}

I would like to thank J. Harvey, E. Martinec, R. Plesser, N. Seiberg
and especially A. Schwimmer
for discussions, and the Physics Departments of Rutgers
University and the Weizmann Institute for hospitality.
This work was supported in part by a DOE OJI award.

\listrefs
\bye